\documentclass[12pt]{article}
\usepackage{amsfonts,bm}
\usepackage{graphicx}
\baselineskip
\offinterlineskip
\begin{document}

\begin{center}
{\bf Hierarchy of Spin Operators, Quantum Gates, Entanglement, Tensor Product and Eigenvalues}
\end{center}

\begin{center}
{\bf Willi-Hans Steeb$^\dag$ and Yorick Hardy$^*$} \\[2ex]

$^\dag$
International School for Scientific Computing, \\
University of Johannesburg, Auckland Park 2006, South Africa, \\
e-mail: {\tt steebwilli@gmail.com}\\[2ex]

$^*$
School of Mathematics, University of the Witwatersrand, \\
Johannesburg, Private Bag 3, Wits 2050, South Africa, \\
email: {\tt yorick.hardy@wits.ac.za}\\[2ex]   
\end{center}

\strut\hfill

{\bf Abstract.} We show that two hierarchies of spin 
Hamilton operators admit the same spectrum. 
Both Hamilton operators play a central role for
quantum gates in particular for the case spin-$\frac12$ and
the case spin-1. The spin-$\frac12$, 
spin-1, spin-$\frac32$ and spin-2 cases are studied in detail.
Entanglement and mutually unbiased bases of the eigenvectors is discussed.
Two triple spin Hamilton operators are also investigated.
Both are also admitting the same spectrum.

\strut\hfill

Let $S_1$, $S_2$, $S_3$ be the spin matrices for spin 
$$
s = \frac12, \,\, 1, \,\, \frac32, \,\, 2,\dots\,\,.
$$ 
The matrices are $(2s+1) \times (2s+1)$ hermitian matrices 
($S_1$ and $S_3$ are real symmetric) with trace equal to 0 satisfying 
the commutation relations
$$
[S_1,S_2] = iS_3, \quad [S_2,S_3] = iS_1, \quad
[S_3,S_1] = iS_2.
$$
As a consequence we obtain $(z \in {\mathbb C})$
\begin{eqnarray*}
e^{zS_1}S_2 e^{-zS_1} &=& \cosh(z)S_2 + i\sinh(z)S_3 \\
e^{zS_2}S_3 e^{-zS_2} &=& \cosh(z)S_3 + i\sinh(z)S_1 \\
e^{zS_3}S_1 e^{-zS_3} &=& \cosh(z)S_1 + i\sinh(z)S_2.
\end{eqnarray*}
The eigenvalues of $S_1$, $S_2$, $S_3$ are $s,s-1,\dots,-s$ for 
a given $s$. It is well known that 
$$
S_1^2 + S_2^2 + S_3^2 = s(s+1)I_{2s+1}
\eqno(2)
$$
where $I_{2s+1}$ is the $(2s+1) \times (2s+1)$ identity matrix.
Furthermore we have
$$
\mbox{tr}(S_j^2)=\frac13 s(s+1)(2s+1)
\eqno(3) 
$$
and
$$
\mbox{tr}(S_jS_k)=0 \quad \mbox{for} \,\,\, j \ne k 
\quad \mbox{and} \,\,\, j,k=1,2,3.
\eqno(4)
$$
In general we have that $\mbox{tr}(S_j^n)=0$ if $n$ is odd and 
$\mbox{tr}(S_j^n)=\frac13 s(s+1)(2s+1)$ if $n$ is even. 
From eq.(4) it follows that $\mbox{tr}((S_jS_k) \otimes (S_\ell S_m))=0$
if $j \ne k$ and $\ell \ne m$. Note that for spin-$\frac12$
we have $S_j=\frac12\sigma_j$ $(j=1,2,3)$, where
$\sigma_1$, $\sigma_2$, $\sigma_3$ are the Pauli spin
matrices.
\newline

We consider the two Hamilton operators (hermitian $(2s+1)^2 \times (2s+1)^2$ matrices) 
$$
\widetilde H = \frac{\hat H}{\hbar\omega} = S_1 \otimes S_1 + S_2 \otimes S_2 + S_3 \otimes S_3
$$
$$
\widetilde K = \frac{\hat K}{\hbar\omega} = 
S_1 \otimes S_2 + S_2 \otimes S_3 + S_3 \otimes S_1
$$
and their spectra. The corresponding quantum gates
$U_{\hat H}(\omega t)=\exp(-i\hat Ht/\hbar)$, 
$U_{\hat K}(\omega t)=\exp(-i\hat Kt/\hbar)$
play a central role in quantum computing \cite{1,2,3}
in particular for the case $s=\frac12$ and the case $s=1$.
The Hamilton operators admit the same spectrum for all given $s$, 
i.e. the eigenvalues (including degeneracy) of the two Hamilton operators 
are the same for a given spin $s$.
\newline

This can be seen as follows. Utilizing eq.(2), eq.(3), eq.(4) 
and that $\mbox{tr}(M \otimes N)=\mbox{tr}(M)\mbox{tr}(N)$ for any $n \times n$
matrices $M$, $N$ one finds that
$$
\mbox{tr}(\widetilde H^j) = \mbox{tr}(\widetilde K^j) 
\quad j=1,2,\dots,(2s+1)^2.
$$
This implies that the eigenvalues of $\widetilde H$ and 
$\widetilde K$ are the same including degeneracy. Note that
for any $n \times n$ matrix $A$ with eigenvalues 
$\lambda_1$, \dots, $\lambda_n$ we have
$$
\mbox{tr}(A^k) = \lambda_1^k + \lambda_2^k + \cdots +
\lambda_n^k
$$
with $k=1,\dots,n$ \cite{4}.
\newline

Consider the cases spin $\frac12$, $1$, $\frac32$ and $2$.
For spin-$\frac12$ we have the hermitian matrices
$$
\widetilde H = \frac14 \pmatrix { 1 & 0 & 0 & 0 \cr 0 & -1 & 2 & 0 \cr
0 & 2 & -1 & 0 \cr 0 & 0 & 0 & 1 } \equiv
\frac14 (1) \oplus \pmatrix { -1 & 2 \cr 2 & -1 } \oplus (1) 
$$
and 
$$
\widetilde K = 
\frac14 \pmatrix { 0 & 1 & -i & -i \cr 1 & 0 & i & i \cr
                   i & -i & 0 & -1 \cr i & -i & -1 & 0 }
$$
with the eigenvalues $-3/4$ (1 $\times$) and $1/4$ (3 $\times$).
Thus the matrices are invertible.
Normalized eigenvectors for $\widetilde H$ are 
$$
\frac1{\sqrt2} \pmatrix { 0 \cr 1 \cr -1 \cr 0 }, \qquad
\pmatrix { 1 \cr 0 \cr 0 \cr 0 }, \,\,\,
\frac1{\sqrt2} \pmatrix { 0 \cr 1 \cr 1 \cr 0 }, \,\,\,
\pmatrix { 0 \cr 0 \cr 0 \cr 1 }.
$$
The first eigenvector (one of the Bell states) is entangled.
The third eigenvector (one of the Bell states) is also
entangled. The second and fourth eigenvectors are not entangled.
However owing to the degeneracy of the eigenvalue $1/4$
we can form two linear combinations of the eigenvectors which provide the 
two remaining Bell states.
\newline

For the matrix $\widetilde K$ normalized eigenvectors are
$$
\frac12 \pmatrix { 1 \cr -1 \cr -i \cr -i }, \qquad
\frac1{\sqrt2} \pmatrix { 1 \cr 0 \cr 0 \cr i }, \,\,\,
\frac1{\sqrt2} \pmatrix { 0 \cr 1 \cr 0 \cr -i }, \,\,\,
\frac1{\sqrt2} \pmatrix { 0 \cr 0 \cr 1 \cr -1 }.
$$
All four vectors are entangled, i.e. none of them
can be written as a Kronecker product of two
vectors in ${\mathbb C}^2$. They form a basis in ${\mathbb C}^4$,
but not an orthonormal basis. Owing to the degeneracy
of the eigenvalue $1/4$ we can always form an orthonormal 
basis in ${\mathbb C}^4$ which are also normalized eigenvectors
$$
\frac12 \pmatrix { 1 \cr -1 \cr -i \cr -i }, \qquad
\frac1{\sqrt2} \pmatrix { 1 \cr 1 \cr 0 \cr 0 }, \,\,\,
\frac12 \pmatrix { 1 \cr -1 \cr i \cr i }, \,\,\,
\frac1{\sqrt2} \pmatrix { 0 \cr 0 \cr 1 \cr -1 }.
$$
Note that the second and fourth eigenvector are not entangled.
This orthonormal basis and the Bell basis
$$
\frac1{\sqrt2}\pmatrix { 0 \cr 1 \cr -1 \cr 0 }, \qquad
\frac1{\sqrt2}\pmatrix { 1 \cr 0 \cr 0 \cr 1 }, \,\,\,
\frac1{\sqrt2}\pmatrix { 0 \cr 1 \cr 1 \cr 0 }, \,\,\,
\frac1{\sqrt2}\pmatrix { 1 \cr 0 \cr 0 \cr -1 }
$$
are mutually unbiased bases \cite{5}. For spin-1 ($9 \times 9$ matrix) 
we find the eigenvalues 
$$
-2 \,\, (1 \times) \,\,\,\,\, -1 \,\, (3 \times), \,\,\,\, +1 \,\, (5 \times).
$$
For spin-$\frac32$ we have $16 \times 16$ matrices and find the eigenvalues
$$
-15/4 \,\, (1 \times), \,\,\,\,\, -11/4 \,\, (3 \times), \,\, 
-3/4 \,\, (5 \times), \,\,\,\, 9/4 \,\, (1 \times).
$$
For spin-2 we have $25 \times 25$ matrices and find the eigenvalues
$$
-6 \,\, (1 \times), \,\,\,\, -5 \,\, (3 \times), \,\,\,\,
-3 \,\, (5 \times), \,\,\,\, 0 \,\, (7 \times), \,\,\,\,
+4 \,\, (9 \times).
$$
The two operators $\widetilde H$ and $\widetilde K$ are connected
by the following unitary matrix $U$. Let
$$
\widetilde H = \sum_{j=0}^{2s}\lambda_j |h_j\rangle\langle h_j|, \quad 
\widetilde K = \sum_{k=0}^{2s}\lambda_k |k_j\rangle\langle k_j|
$$
be the spectral decomposition of $\widetilde H$ and $\widetilde K$,
respectively. Then with the unitary matrix
$$
U = \sum_{\ell=0}^{2s} |k_{\ell}\rangle\langle h_{\ell}|
$$
we have $U\widetilde H U^*=\widetilde K$.
\newline

We also note that $[S_1 \otimes S_1,S_2 \otimes S_2]=0_2 \otimes 0_2$
for spin-$\frac12$. However for spin-1 we have
$[S_1 \otimes S_1,S_2 \otimes S_2] \ne 0_2 \otimes 0_2$.
\newline

For $s=1/2$ the unitary matrix (quantum gate) $U(t)=\exp(-i\hat Ht/\hbar)$
is given by
$$
U(t) =
(e^{-i\omega t/4}) \oplus
e^{it\omega/4}\pmatrix{
\cos(\omega t/2) & -i\sin(\omega t/2) \cr
-i\sin(\omega t/2) & \cos(\omega t/2) } \oplus (e^{-i\omega t/4})
$$
where $\oplus$ denotes the direct sum.
For $s=1/2$ the unitary matrix (quantum gate) $V(t)=\exp(-i\hat Kt/\hbar)$
is given by
$$
V(t) =
\frac{e^{i\omega t/4}}2
\pmatrix{ \cos(\omega t/2) & -i\sin(\omega t/2) & -\sin(\omega t/2) & -\sin(\omega t/2) \cr
-i\sin(\omega t/2) & \cos(\omega t/2) & \sin(\omega t/2) & \sin(\omega t/2) \cr
\sin(\omega t/2) & -\sin(\omega t/2) & \cos(\omega t/2) & i\sin(\omega t/2) \cr
\sin(\omega t/2) & -\sin(\omega t/2) & i\sin(\omega t/2) & \cos(\omega t/2) }
+\frac{e^{-i\omega t/4}}{2}I_4.
$$
Consider now the triple spin operators \cite{6} for spin-$\frac12$
$$
\widetilde H = \frac{\hat H}{\hbar\omega} =
S_1 \otimes S_1 \otimes S_1 + S_2 \otimes S_2 \otimes S_2 +
S_3 \otimes S_3 \otimes S_3 
$$
and
$$
\widetilde K = \frac{\hat K}{\hbar\omega} =
S_1 \otimes S_2 \otimes S_3 + S_3 \otimes S_1 \otimes S_2 +
S_2 \otimes S_3 \otimes S_1. 
$$
Both admit the same spectrum given by
$$
-\frac18\sqrt3 \,\, (4 \, \times), \qquad
\frac18\sqrt3 \,\, (4 \,\times). 
$$
For higher order spin we also find the same spectrum 
for the two triple spin Hamilton operators.

\end{document}